\documentstyle[12pt,twoside]{article}
\begin{document}
\title
{SQUARED WEYL AND DIRAC FIELDS WITH SOMMERS -- SEN CONNECTION, ASSOCIATED
WITH  $V^3_4$ DISTRIBUTION}

  \author{Volodymyr Pelykh}

\maketitle

\begin{abstract}
The generalization of Sommers--Sen spinor connection for spinor fields,
associated with a distribution $V_4^3$ and on this basis the equations for
Weyl and Dirac null vector fields on complexificated $V_4^3$ are obtained.
 We interpret the obtained results by examining the interaction of spinor
fields with the inertial forces.
\end{abstract}

\section{INTRODUCTION}

As known, the four--dimensional description of relativistic fields for the row
of problems, especially under the comparison of theoretical previsions and
experimental results, must by exchanged by 3+1 description.
For the spinor fields the original method of their description in 3+1 form
was proposed in \cite{Som} and developed in \cite{Sen}.
The obtained in \cite{Sen} Sen--Witten equation has a wide usage for the
positive signification of gravitational field energy problem investigations
\cite{Ash,Reu,Frau}. But this method does not allow to study all the variety
of physical effects in the interactions of spinor fields with inertial forces,
because in the base of the method lies the foliation of curved space-time
by space-like hypersurfaces and thus the nonrotatory frames of reference.
In our work \cite{Pel} we introduced a covariant derivative of spinor fields,
associated with the space-like distribution $V^3_4$, which generalizes the
Sommers--Sen
covariant derivative, and obtained on this base the  3+1 equation for Weyl and
Rarita--Schwinger fields in
arbitrary frames of reference, not only in nonrotatory.
In this work we obtain for arbitrary frame of reference the equations for
complex 3-vector, which are correspondent to Weyl spinors and bispinors.
This squared equtions of Weyl and Dirac field are in fact the 3+1 splitting
of Penrose--Rindler \cite{Pen} tensor form of spinor differential equations.

 In section 2 we briefly review the technique for obtaining the 3+1 spinor
equations with spinor connection on nonintegrable manifolds, in section 3 we
give the spinor representation of tensor fields on the $V_4^3$ distribution.

 \section{THE SOMMERS--SEN GENERALIZED DERIVATIVE ON DISTRIBUTIONS}

 Let us consider the oriented manifold $V_4$ of class $C^{\infty},$ which is
 noncompact or of zero Euler characteristic; then it accepts Lorentz
  $\bf g$ metric.
 Let us denote, as usually, the tangent bundle over $V_4$ by $TV_4.$

       {\bf Definition 1.} A vector subbundle $V_4^m,\quad 1\le
   m\le 3,$ of $TV_4$ is called a distribution over $V_4$.
   Let us denote by $C^{\infty}(V_4)$ a ring of functions of
$C^{\infty}$ class and by $\Gamma(V_4^m)$ --- a \, $C^{\infty}$ module of
sections of distribution $V_4^m$ over $V_4$.

            {\bf Definition 2.} The mapping
 $$
    \Gamma(V_4^m) \ \times \ \Gamma(V_4^m) \
   \longrightarrow \ \Gamma(V_4^m),$$
    $$({\bf X},{\bf X}_1) \
   \longrightarrow  { D}_{{\bf X}}{\bf X}_1,\qquad {\bf
   X}, {\bf X_1} \in \Gamma(V_4^m),
$$
which for arbitrary vectors
 ${\bf
X}, {\bf X}_1$ and functions $g, f \in C^\infty (V_4)$ satisfies
the conditions

   \begin{equation}
   \label{1} { D}_{{\bf X}} ({\bf X_{1}}+{\bf X}_{2})  =
   { D}_{{\bf X}}{\bf X}_1
   +{ D}_{{\bf X}}{\bf X}_{2},
    \end{equation}
   \begin{equation} \label{2} { D}_{{\bf X}}(f{\bf
    X_{1})} = {\bf X}(f){\bf X}_{1}+f{ D}_{{\bf  X}} {\bf
    X}_{1}, \end{equation} \begin{equation} \label{3}
    {D}_{f{\bf X}+g{\bf X}_{1}}{\bf X}_{2}  =
   f{ D}_{\bf X} {\bf X}_{2}+g{ D}_{{{\bf
   X}_{1}}}{\bf X}_{2}.  \end{equation}
is called a covariant derivative or anholonomic connection on distribution
$V_4^m$

The homomorphism
 $$
    \Gamma (B_{0}):\quad \Gamma (TV_4)  \longrightarrow  \Gamma (V_4^m)
$$
 of cross-sections is corresponding with
the projections  $B_{0}$ of bundle  $TV_4$ on subbundle  $V_4^m.$

   {\bf Definition 3.} $im\: C_0,$ where $C_0=1_{TV_4}-B_0,$ is called a rigging
$V_4^{4-m}$   of distribution $V_4^m\subset TV_4.$

Let $V_4^1$ be a one-dimensional time-like distribution over $V_4$;
   it's unitary cross-section {\bf u} is identified with the field of
   4-velocity of some frame of reference, and the integral curve of cross-section
   {\bf u} --- with is time lines. The normal rigging $V_4^3$
   of the distribution $V_4^1$ is the geometrical image of the physical space for
appropriate frame of reference  and is nonintegrable in general.
   The vectors, which belong to $V_4^1,$ are called the time vectors, and
   those, which belong to $V_4^3$ ---the spatial vectors.

   Further we require that the second   Stiefel--Whitney class $w_2$
of manifold
   $V_4$ equals zero. Then $V_4$ permits the $SL(2,C)\!$ spinor structure.
   Let us denote by $ {\cal S}_{\,r,s}(V_4)$ the $C^{\infty}\!$ module
   of spinor fields of ($r,s$) valence on $V_4.$

   Let us introduce $C^{\infty}\!$ module of  ${\cal S}_{\,r,s}(V_4^3)$
   spinor fields of ($r,\,s$) valence on $V_4,$ associated with the distribution
   $V_4^3\,,$ in the next way: its elements are the spinor fields of the form
    $\varphi T^{A\ldots L}{}_{M\ldots Q}$ and
   \begin{eqnarray*}
   \psi T^{A\ldots KC\ldots L}{}_{M\ldots RP\ldots Q}&\!\!\!=&\!\!\!\left(-\sqrt {2}
   u^A_{\dot A}\right)\;\ldots\;\left(-\sqrt {2}u_{\dot K}^K\right)
   \left(\sqrt {2}u^{\dot M}_M\right)
   \;\ldots\;\left(\sqrt{2}u^{\dot R}_R\right)\times \\[1ex]
   &\!\!\!&\!\!\!
   T^{{\dot A}\ldots\dot KC\ldots L}{}_{\dot M\ldots\dot RP\ldots Q}\,,
   \end{eqnarray*}
   where $\varphi\, $ and $\psi \,$ are the arbitrary $C^{\infty}(V_4)$ functions.
   In such way defined the $C^{\infty}\!$ module ${\cal S}_{r,s}(V_4^3)$ is the
   module of $SU(2)\!$ spinor fields. A module  ${\cal S}_{r,s}(V_4^3)$ with
    the basis limited to the hypersurface $\Sigma$ let call
    a module of $SU(2)$ spinor fields on the anholonomic hypersurface $\Sigma$.
    In the partial case, when this hypersurface is ordinary and space--like,
   the module of $SU(2)$ spinor fields on it coincides with the module of
     Sommers--Sen spinor fields.

   We introduce the antisymmetric tensor ${\bf A}$
anholonomicity
of  $V^3_4 ,\;{\bf A}\in V_4^3$.   Let
$$ {\bf T}\, :\Gamma(V_4^3)  \times \ \Gamma(V_4^3)  \longrightarrow
 \Gamma(V_4^1),$$
$${\bf T}=\Gamma (C_0)\,[{\bf X}_1,{\bf X}_2];\qquad {\bf X_1},\;{\bf X_2}\;
\in\Gamma(V_4^3).$$
Then ${\bf T}=4{\bf A}\otimes{\bf u}$. In coordinate basis on some open
domain
in $V_4$ the tensor ${\bf A}$ has the components
\footnote{For expressions written in coordinates, Greek indices are global
and Latin indices local. Our Lorentz metric will have signature (-2)}

$$A_{\mu\lambda}=\frac{1}{2}h_\mu^\nu h_\lambda^\delta \nabla_{[\nu}u_{\delta]}\,.$$

Let us introduce the spatial covariant derivative for spinor fields, associated
with the distribution $V_4^3,$ as the mapping
$$\Gamma(V_4^3)\times{\cal S}_{\,1,0}\,(V_4^3)\,\longrightarrow \,
{\cal S}_{\,1,0}\,(V_4^3),$$
\begin{equation}\label{2}({\bf X},\,\lambda)\,\longrightarrow \,
{ D}_{{\bf X}}\lambda,\quad
{\bf X}\,\in V_4^3\,, \quad \lambda \in {\cal S}_{\,1,0}\,(V_4^3)
\end{equation}
determined by the condition
\begin{equation}\label{3}{ D}_{AB}\lambda_C=
\sqrt{2}u_{(A}{}^{\dot A}\nabla_{B)\dot A}\lambda_C-
\frac{1}{\sqrt{2}}(\pi_{ABC}{}^D+A_{ABC}{}^D)\lambda_D,
\end{equation}
where $\nabla_{B\dot A}$ is the spinor representation of the operator of
covariant derivative on $V_4,$ in agreement with metrical connection. The
action ${ D}_{AB}$ on spinors of higher valence extends in accordance
with the Leibnitz rule, and the action on vector fields satisfies the
condition (\ref{1})--(\ref{3}).

Reducing the $SL(2,C)$ operator of covariant derivative to $SU(2)$
operator, we obtain:
\bigskip
$$\nabla_{AB}= \sqrt{2}\left(
u^{\dot A}{}_{[B}\nabla_{A]{\dot A}}+u^{\dot A}{}_{(B}\nabla_{A){\dot A}}
\right)=\frac{\sqrt2}{2}\varepsilon_{AB}u^{A{\dot A}}\nabla_{A{\dot A}}+
\sqrt{2}n^{\dot A}{}_{(B}\nabla_{A){\dot A}}\,.
$$
The first term, denoted by $\frac{\sqrt2}{2}\varepsilon_{AB}
({\bf u\cdot\nabla}),$ is the time derivative, the second is represented
in terms space derivative ${ D}_{AB}$ in the rigging.
Finally, we obtain the action of $\nabla_{AB}$ on spinor $\lambda_C$ in form
\begin{equation}\label{4}
\nabla_{AB}\lambda_C=\frac{\sqrt{2}}{2}\varepsilon_{AB}u^{A{\dot A}}
\nabla_{A{\dot A}}\lambda_C+{ D}_{AB}\lambda_C-\frac{\sqrt{2}}{2}\left(
\pi_{AB}{}^C{}_D+
A_{AB}{}^C{}_D\right)\lambda^D .
\end{equation}

 The generalized 3+1 form of Weyl equation

$$
      \nabla_{A{\dot A}}\lambda^A=0
$$
 we obtain, fulfilling the  $SL(2,C)\,\rightarrow\, SU(2)$
reduction and using (\ref{4}). Then we have:
\begin{equation}\label{...}
({\bf u\cdot\nabla})\lambda _A +\sqrt{2}{ D}_{AB}\lambda^B+
\frac12\pi\lambda_A-
A_{BA}{}^B{}_D\lambda^D=0.
\end{equation}
Therefore the Weyl spinor $\lambda_A\in {\cal S}_{1,0}(V_4^3)$ is
determined by both geometric and equally
physical properties of $V_4^3$ and $V_4^1$. These properties are determined by
the acceleration spinor $F_{AB}\in{\cal S}_{\,2,0}(V_4^3)$,
the angular velocity spinor $A_{ABCD}\in{\cal S}_{\,4,0}(V_4^3)$ and
the rate--of--strain spinor $\pi_{ABCD}\in{\cal S}_{\,4,0}(V_4^3)$.
These spinors are uniquely expressed by the Schouten first order curvature
tensors of $V_4^3$ and vector ${\bf u}\in V_4^1$\,.

\section{SPINOR REPRESENTATION OF TENSORS FIELDS ON THE  $V^3_4$ DISTRIBUTIONS}

Let  spinor field ${\bf T}\in S_{2r,2s}(V^3_4)$. If  ${\bf T}$ is symmetric
in all pairs of indices, then ${\bf T}\in V^3_4$. The projector from $TV_4$
into $V^1_4$ is ${\bf  u}\otimes {\bf  u}$, the projector from $TV_4$
in $V^3_4$ is ${\bf h}={\bf g}-{\bf u}\otimes {\bf  u}$.

It is easy to characterize the $SU(2)$ representation of a spatial projected
tensor:


$$\overline{v}_{AB}=\overline{v}_{A{\dot A}}{\sqrt{2}}u_B{}^{\dot A}=
 v_{\nu}(g^\nu_\mu-u^\nu u_\mu)h^\mu_m\sigma^m_{A{\dot A}}=
 v_\nu(g^\nu_\mu-u^\nu u_\mu)\sigma^\mu_{A{\dot A}}=$$
$$
 \overline{v}_{\mu}\sigma^\mu{}_{A{\dot A}}{\sqrt{2}}u_B{}^{\dot A}=
\overline{v}_m\sigma^m_{AB}=\overline{v}_\mu\sigma^\mu_{AB}\,.$$

Overline denotes the components of spatial projected tensors, the symbol
\^{} denotes the components of time projected tensors, $\sqrt2\sigma^\mu_{A{\dot A}}$
are the Pauli spin matrices and the unit matrix which are referred to a
space--time
tetrad. The    $\sigma^\lambda_{AB}$ and $\sigma^l_{AB}$  matrices are given
by the formulas
$$\sigma^l_{AB}=\sigma^l_{A{\dot A}}{\sqrt{2}}u_B{}^{\dot A}=
 {\sqrt{2}}u^m\sigma^l_{A{\dot A}}\sigma_{mB}{}^{\dot A} $$

and respectively
$$\sigma^\lambda{}_{AB}=\sigma^\lambda{}_{A{\dot A}} {\sqrt{2}}u_B{}^{\dot A}=
{\sqrt{2}}u^\mu\sigma^\lambda{}_{A{\dot A}}\sigma_{\mu B}{}^{\dot A}\,.
$$
For the matrices    $\sigma^\lambda_{AB}$ we obtain necessary in the following
consideration identities, which exchange the normalization and ortogonalization
identities of Pauli matrices:
$$\sigma^\lambda{}_{AB}\sigma_\lambda{}^{DC}=\sigma^l{}_{AB}\sigma_l{}^{DC}=
{\sqrt{2}}u^\mu\sigma_{\mu B}{}^{{\dot A}}\sigma^\lambda{}_{A{\dot A}}
{\sqrt{2}}u^\nu\sigma_\nu{}^{C{\dot C}}\sigma_{\lambda}{}^D{}_{\dot C}=
\sqrt{2}u^\nu\sigma_\nu{}^C{}_{B}\delta_A^D=$$
$$
\sqrt{2}\widehat{\sigma}^C{}_{B}\delta^D_A      \,,
\qquad
\sigma^\lambda{}_{AB}\sigma_\mu{}^{AB}=\delta_\mu^\lambda\,.
$$
Let us also obtain the necessary for the further investigations
tensor representation of
the spinor form
\begin{equation}\label{8}
 \overline  y_{C(A}\overline z_{B)}{}^C=
 \sum_{m=0}^3 \overline  y_m\sigma^m_{C(A}
\overline z_m\sigma^m{}_{B)}{}^C+
 \sum_{m\ne l}^3 \overline  y_m\overline z_l\sigma^m_{C(A}
\sigma^l{}_{B)}{}^C.
\end{equation}
The matrices $\sigma^m_{CA}$ we obtain in the form
$$\sigma^m{}_{CA}=\sqrt2(n^0\omega^m{}_{CA}+n^1\tau^m{}_{CA}+n^2\chi^m{}_{CA}+
n^3\psi^m{}_{CA}),\quad\sigma^l{}_{C}{}^B=\sigma^l{}_{CA}\varepsilon^{BC},$$
where
$$\omega^m{}_{CA}=\sigma^m{}_{C\dot C}\sigma_{0A}{}^{\dot C},\;
\tau^m{}_{CA}=\sigma^m{}_{C\dot C}\sigma_{1A}{}^{\dot C},\;
\chi^m{}_{CA}=\sigma^m{}_{C\dot C}\sigma_{2A}{}^{\dot C},\;
$$
$$
\psi^m{}_{CA}=\sigma^m{}_{C\dot C}\sigma_{3A}{}^{\dot C}\,.
$$
The first sum in  (\ref{8}) equals zero; therefore further we consider only the second sum,
in which we distinguish  the terms with the products
$\overline  y_{1}\overline z_{2}$. In this case the products of necessary
matrices give:
$$
\begin{array}{ll}
\omega^1{}_{CA}\omega^2{}_{B}{}^C=-i\omega^3{}_{AB}
& \hspace{1cm}\tau^1{}_{CA}\omega^2{}_{B}{}^C=-i\tau^3{}_{AB}\\
\omega^1{}_{CA}\tau^2{}_{B}{}^C=\chi^1{}_{AB}
& \hspace{1cm}\tau^1{}_{CA}\tau^2{}_{B}{}^C=i\tau^3{}_{AB}\\
\omega^1{}_{CA}\chi^2{}_{B}{}^C=-i\chi^3{}_{AB}
& \hspace{1cm}\tau^1{}_{CA}\chi^2{}_{B}{}^C=-\chi^2{}_{AB}\\
\omega^1{}_{CA}\psi^2{}_{B}{}^C=-i\psi^3{}_{AB}
& \hspace{1cm}\tau^1{}_{CA}\psi^2{}_{B}{}^C=-\psi^2{}_{AB}
\end{array}
$$
$$
\vspace{1cm}
\begin{array}{ll}
\chi^1{}_{CA}\omega^2{}_{B}{}^C=-i\chi^3{}_{AB}
& \hspace{1cm}\psi^1{}_{CA}\omega^2{}_{B}{}^C=\omega^0{}_{AB}\\
\chi^1{}_{CA}\tau^2{}_{B}{}^C=-\tau^1{}_{AB}
& \hspace{1cm}\psi^1{}_{CA}\tau^2{}_{B}{}^C=i\tau^0{}_{AB}\\
\chi^1{}_{CA}\chi^2{}_{B}{}^C=i\psi^0{}_{AB}
& \hspace{1cm}\psi^1{}_{CA}\chi^2{}_{B}{}^C=-i\psi^1{}_{AB}\\
\chi^1{}_{CA}\psi^2{}_{B}{}^C=-\psi^2{}_{AB}
& \hspace{1cm}\psi^1{}_{CA}\psi^2{}_{B}{}^C=i\psi^0{}_{AB}\,.
\end{array}
$$
After this the term with
$\overline  y_{1}\overline z_{2}$ we obtain in the form:
$$
\sqrt2\overline  y_{1}\overline z_{2}[-in^0\sigma^3_{(AB)}+
in^3\sigma^0_{(AB)}]\,.
$$
By analogy for terms with  $\overline  y_{2}\overline z_{1}$ we have
$$
\sqrt2\overline  y_{2}\overline z_{1}[in^0\sigma^3_{(AB)}-in^3\sigma^0_{(AB)}]\, .
$$
After summarizing of all terms of the form
$
\overline  y_{m}\overline z_{l}\quad(m\ne l)
$
 we find, that
\begin{equation}\label{9}
\overline  y_{C(A}\overline z_{B)}{}^C=-i\sqrt2\varepsilon^{asn}{}_kn_a
\overline  y_{s}\overline z_{n}\sigma^k_{(AB)}.
\end{equation}
Let us consider the evolution of complex null spatial vector field
${\bf L}=-\lambda_A\lambda_B,$
corresponding to the Weyl spinor field. Then as first step we obtain
$$
({\bf u}\cdot\nabla)L_a=2\sqrt2\lambda_{(B}D_{A)C}\lambda^C-\pi L_a+
2\lambda_{(B}A_{A)C}{}^C{}_D\lambda^D.
$$
After the using of the Sommers identity
$$
\lambda^CD_{AB}\lambda_C=\lambda^CD_{C(A}\lambda_{B)}-
\lambda_{(A}D_{B)C}\lambda^C
$$
we have
$$
({\bf u}\cdot\nabla)L_a=-\sqrt2\lambda^CD_{AB}\lambda_C-\pi L_a
-\sqrt2D_{C(A}L_{B)}{}^C+2\lambda_{(B}A_{A)C}{}^C{}_D\lambda^D.
$$
For the vector representation of $ \sqrt2D_{C(A}L_{B)}{}^C$ we obtain:
$$
D_{C(A}L_{B)}{}^C=-i\sqrt2\varepsilon^{asn}{}_kn_a\sigma^k{}_{(AB)}D_sL_n-
i\sqrt2\varepsilon^{asn}{}_kA_{as}L_n
\sigma^k{}_{(AB)}=
$$
$$
-\sqrt2\cdot3!i\,*({\bf A}\land{\bf L}+
{\bf u}\land {\bf D}\land{\bf L}).
$$
The first term on the left can be re--expressed \cite{Som} as
$\widetilde{{\bf L}}^cD_aL_c$, where unit real spatial vector
$\widetilde{{\bf L}}$ is in the direction of propagation of
 the neutrino field and is
$$
\widetilde{L}^c=-i(L_d\bar L^d)^{-1} \varepsilon^a{}_{sn}{}^{c}n_a
L^s\overline{L}^n.
$$
Finally, we find that squared neutrino equation on $V_4^1$ and $V_4^3$
distributions,
i.e. in a arbitrary frame of reference is
$$
<{\bf dL,u}>=<\widetilde{{\bf L}},{\bf DL}>-\pi{\bf L}+
2\cdot3!\,i*({\bf u}\land {\bf D}\land{\bf L})+
$$
$$
4\cdot3!*({\bf u}\land {\bf w}\land{\bf L})-
2\cdot3!\,i*({\bf u}\land {\bf F}\land{\bf L})-
\sqrt2\cdot3!i\,*({\bf A}\land{\bf L}).
$$
  With the help of squared Weyl equations we can simply obtain the squared
Dirac equation in 3+1 form.
The Dirac equation is equivalent  to the pair of equations
$$
\nabla_{A\dot A}\xi^A=\frac m{\sqrt2}\eta_{\dot A}\hspace{1cm}\mbox
{and}\hspace{1cm} \nabla^{A\dot A}
 \eta_{\dot A}=-\frac m{\sqrt2}\xi^A\,
$$
or in terms of\, $\xi^A\in {\cal S}_{\,0,1}(V_4^3),\:
\eta_B\in {\cal S}_{\,1,0}(V_4^3)\,:$
$$
\nabla_{AB}\xi^A=\frac m{\sqrt2}\eta_B\hspace{1cm}\mbox{and}\hspace{1cm}\nabla^{AB}\eta_B =
-\frac m{\sqrt2}\xi^A\,.
$$
Let ${\bf X}=-\xi_A\xi_B$ and ${\bf Y}=-\eta_A\eta_B$ are two complex null
spatial
vector fields. Then the squared Dirac equation, which describes the evolution
of    ${\bf X}$ and ${\bf Y}$ fields and which is written in terms
of the tensors
determined on the distribution and in the rigging, is the system of
two equations:
$$
<{\bf dX,u}>=<\widetilde{{\bf X}},{\bf DX}> -\pi{\bf X}
-2\cdot3!\,i*\left[{\bf u}\land ({\bf F}-{\bf D}+2i{\bf w})\land{\bf X}-
{\bf A}\land{\bf X}\right]-
$$
$$
3!\,\sqrt2im<{-\bf X,\;Y}>^{-1/2}*({\bf u}\land {\bf X}\land{\bf Y})
$$
and
$$
<{\bf dY,u}>=-<\widetilde{{\bf Y}},{\bf DY}> -\pi{\bf Y}
+2\cdot3!\,i*\left[{\bf u}\land ({\bf F}+{\bf D}-2i{\bf w})\land{\bf Y}+
{\bf A}\land{\bf Y}\right]-
$$
$$
3!\,\sqrt2im<{-\bf X,\;Y}>^{-1/2}*({\bf u}\land {\bf X}\land{\bf Y}).
$$
\section{DISCUSSION}

The proposed in \cite{Pel} and in this paper method for investigation of the
interaction between spinor fields and inertial forces requires the usage of
nonintegrable subbundle. In difference to \cite{Som,Sen} the spinor
derivatives are determined here by intrinsic geometry of distribution.
This is defined by the physical sense of the problem and not by
application of tetrad formalism, which was not necessary. In partial case of
integrable $V_4^3$ distribution both tetrad and monad methods determine the
spinors in terms of the intrinsic geometry of foliation.

We ascertain the appearance of additional difference between the evolution
of Maxwell field, Weyl field and Dirac   field, since the interaction of latter
with the inertial field is described by the term, which includes not
only the angular velocity vector of the frame of reference ${\bf w}$\,, but also
its angular velocity tensor ${\bf A}$\,.

\end{document}